\documentstyle[11pt]{article}
\date{}

\begin{document}
\title{Quark and Gluon Propagators from Meson Data}
\author{{Reginald T. Cahill  and      Susan M. Gunner
  \thanks{E-mail: Reg.Cahill@flinders.edu.au,
ECSMG@psy1.ssn.flinders.edu.au}}\\
  {Department of Physics, Flinders University}\\ { GPO Box 2100, Adelaide 5001,
Australia }\\
{June 1995}}

\maketitle

\begin{center}
\begin{minipage}{120mm}
\vskip 0.6in
\begin{center}{\bf Abstract}\end{center}
{We report  robust  calculations of various low energy QCD
hadronic properties. We use a multi-rank separable expansion for the gluon
propagator
which greatly facilitates the numerical computations within the Global Colour
Model for QCD. The parameters for the  propagators are
determined by fitting  experimental values for $f_{\pi}$ and the  $\pi$ and
$a_1$
meson masses.

}
\end{minipage} \end{center}

\vskip 1.5cm
\par

The computation of the low energy properties of QCD is a difficult
non-perturbative problem
in quantum field theory. Here we report new results using the Global Colour
Model (GCM)
\cite{CR85} approximation to QCD. These  robust computations are made feasible
by
the use of multi-rank separable expansions of the gluon propagator.  While good
progress \cite{Pennington} has been made in computing the gluon propagator from
first
principles in QCD, here we adopt the procedure of determining an effective
gluon propagator
by using a convenient separable form, and selecting the parameters therein by
fitting
$f_{\pi}$ and the $\pi$ and  $a_1$ meson masses to experimental values.  The
validity of both
the GCM and the effective separable-form gluon propagator are  demonstrated by
then
computing numerous other low energy hadronic properties.  A non-separable
translation-invariant  form for the effective gluon propagator, appropriate to
low energy
hadronic states, is then reconstructed from the separable form.

An overview and an insight into the nature of the non-perturbative low energy
hadronic regime
of  QCD is provided by the functional integral hadronization of QCD
\cite{RTC,Reinhardt90}.
This amounts to a dynamically determined change of functional integration
variables, from
quarks and gluons, to bare hadrons

$
\int {\cal D}\overline{q}{\cal D}q{\cal
D}Aexp(-S_{qcd}[A,\overline{q},q]+\overline{\eta}q+ \overline{q}\eta)\approx$
\begin{equation} \int{\cal
D}\pi{\cal D}\overline{N}{\cal
D}N...exp(-S_{had}[\pi,...,\overline{N},N,..]+J_{\pi}[\overline{\eta},\eta]\pi+..)
\end{equation}

  The final functional integration over the hadrons  gives the
hadronic observables, and amounts to dressing each hadron by, mainly, lighter
mesons.  This
functional integral transformation cannot  yet be done exactly.  The basic
insight is that the
quark-gluon dynamics, on the LHS of (1),  is fluctuation dominated, whereas the
RHS is not, and
for example the meson dressing of bare hadrons is known to be almost
perturbative. In performing
the change of variables essentially normal mode  techniques are used
\cite{RTC}. In practice this
requires detailed numerical computation of the gluon propagator, quark
propagators, and  meson
and baryon propagators. The mass-shell states of the latter are determined by
covariant
Bethe-Salpeter  and  Faddeev equations. The Faddeev computations are made
feasible by using the
diquark correlation propagators, which must also be determined.

 The first and easiest formal
transformation results from doing the gluon integrations, leaving an action for
quarks of the
form
 \begin{equation}S[\overline{q},q]= \int
\overline{q}(x)(-\gamma . \partial+{\cal M}) q(x) +\frac{1}{2}\int
j^a_{\mu}(x)j^a_{\nu}(y)D_{\mu\nu}(x-y)+\frac{1}{3!}\int
j^a_{\mu}j^b_{\nu}j^c_{\rho}D^{abc}_{\mu\nu\rho}+...
\end{equation}
where $j^a_{\mu}(x)=\overline{q}(x)\frac{\lambda^a}{2}\gamma_{\mu}q(x)$.
The GCM  is a model field theory for QCD  based on a  truncation
of $S[\overline{q},q]$ in which
 the  higher order n-point ($n\geq 3$) functions are neglected, and   only the
gluon 2-point
function  $D_{\mu\nu}(x-y)$  is retained.

The GCM is thus a quantum field theory that can also  be considered to be
defined by the action
$$
S_{gcm}[\overline{q},q,A^a_{\mu}]=\int \left(
\overline{q}(x)(-\gamma . \partial+{\cal
M}+iA^a_{\mu}\frac{\lambda^a}{2}\gamma_{\mu})\delta(x-y)q(y) +\right.$$
\begin{equation}
\left. +\frac{1}{2}
A^a_{\mu}(x)D^{-1}_{\mu\nu}(i\partial)\delta(x-y)A^a_{\nu}(y) \right)
\end{equation}
where the matrix $D^{-1}_{\mu\nu}(p)$  is the inverse of $D_{\mu\nu}(p)$, which
in turn is the
Fourier transform of $D_{\mu\nu}(x)$. This action has a
global colour symmetry. The GCM is thus analogous to QED except for  colour
currents and
the non-quadratic phenomenological form for $D^{-1}_{\mu\nu}(p)$  in the pure
gluon sector.
The main purpose of this work is to report a robust and effective procedure for
determining
$D_{\mu\nu}(p)$ and to demonstrate its general validity for  a range of
hadronic phenomena.
This procedure is to use the separable expansion technique which proved very
effective in the
1960's in studying non-relativistic few particle systems.

Having made the GCM truncation in (2) it is possible to proceed further and
to transform \cite{RTC} the quark functional
integrations into the hadronic functional integrations, as in (1). If the
additional approximation
$D_{\mu\nu}(x-y)\rightarrow g\delta_{\mu\nu}\delta(x-y)$ is made in (2), i.e. a
contact coupling of the quark currents, then the NJL type models are obtained.
If in (1)
a derivative expansion of the complete non-local  hadronic effective action is
performed, then the
Chiral Perturbation Theory (CPT) phenomenology is obtained. However in the GCM,
with appropriate
$D_{\mu\nu}(x)$, all computations are finite and no cutoffs or renormalisation
procedures are used. As
well, using a mean field approximation, the soliton phenomenology for the
baryons may be derived
\cite{CR85}, and has been studied in
\cite{Frank}.

 As shown here and elsewhere this approximation is
surprisingly effective in describing the low energy QCD determined hadronic
properties, and
suggests that some particular mechanism is responsible for its success.
There are indications that the neglected terms in (2) may only be significant
in
those bound states which are not colour singlets \cite{RTC}. In colour singlet
states some colour
neutrality may render the higher order terms ineffective. In the GCM the
remaining gluon 2-point
 function is now regarded as an effective 2-point function:
$D_{\mu\nu}(x-y)\rightarrow
D_{\mu\nu}(x-y)_{eff}$.  There are then two key steps in the GCM: (1) the
determination of this effective 2-point function, and the demonstration that
the same function
does  well in determining a variety of hadronic observables, so that in some
sense it
is {\it universal}, and (2) the comparison of $D_{eff}$ with the true one
obtained
from QCD. This would establish to what extent the effects of the higher order
terms have
been accounted for by $D_{eff}$. From now on we shall call this $D$.

Having specified $ D(p)$ the usual procedure would be to first determine the
(constituent) quark
propagators, in which $m$ is the quark current mass,
\begin{equation}
G(q)=(iA(q)q.\gamma+B(q)+m)^{-1}=-iq.\gamma\sigma_v(q)+\sigma_s(q)
\end{equation}
using the non-linear Dyson-Schwinger equations (DSE), in Euclidean metric,
\begin{equation}
B(p)=\frac{16}{3}\int\frac{d^4q}{(2\pi)^4}D(p-q).\frac{B(q)+m}{q^2A(q)^2+(B(q)+m)^2},
\end{equation}

\begin{equation}
[A(p)-1]p^2=\frac{8}{3}\int\frac{d^4q}{(2\pi)^4}q.pD(p-q).\frac{A(q)}{q^2A(q)^2+(B(q)+m)^2},
\end{equation}
where (5), using the convolution theorem for Fourier
transforms, has a particularly simple form in coordinate space
\begin{equation}
B(x)=\frac{16}{3}D(x)\sigma_s(x).
\end{equation}
Eqn.(5) is the more sensitive one of (5) and (6): the meson form factors are
closely related to
the extent of $B(p)$, whereas $A(p)$ is  changing more slowly. For simplicity
we have used a
Feynman-like gauge, and  the perturbative quark-gluon vertex function. One can
choose to use the
GCM in other gauges. If one chooses a  Landau gauge form
$$D_{\mu\nu}(p)=(\delta_{\mu\nu}-\frac{p_{\mu}p_{\nu}}{p^2})\Delta(p^2)$$ then
from the $O(4)$
invariance of $B(p)$, we obtain   that $D(p^2) \equiv \frac{3}{4}\Delta(p^2)$
in (5). A similar
effect occurs in the BSE because for the low mass states, with the confining
quark propagator, the
form factors are almost O(4) invariant wrt the relative momentum dependence
\cite{Stainsby94}.

The DSE only have the forms in (5) and
(6) for a translation invariant action, as  in (2) or (3). To solve these
equations for various
$D(p)$ and to analytically continue $A(s)$ and $B(s)$ into the complex s-plane,
where $s=q^2$,
when solving for meson, diquark and baryon propagators is particularly
difficult. Hence
we  have   studied the well known separable expansion technique.  This greatly
facilitates the
solutions of the SDE, the BSE and the Faddeev equations.

We first imagine expanding   $D(p-q)$ into $O(4)$ hyperspherical harmonics
\begin{equation}
D(p-q)=D_0(p^2,q^2)+q.pD_1(p^2,q^2)+...
\end{equation}
where, for example,
\begin{equation}
D_0(p^2,q^2)=\frac{2}{\pi}\int_0^{\pi}d\beta\,sin^2\beta\, D(p^2+q^2-2 p q
cos\beta).
\end{equation}
However note that putting $q=0$ in (9) gives  $D(p)=D_0(p^2,0)$. Thus  the full
$D(p)$  may be
easily reconstructed from the first term in the RHS of (8).
 Note also that only the first two terms in (8) are needed in (5) and (6);
higher order
terms do not contribute as they are orthogonal to the measures therein.

To facilitate the
numerical computations we then introduce multi-rank separable expansions for
each term
\begin{equation}
D_0(p^2,q^2)=\sum_{i=1,n} \Gamma_i(p^2)\Gamma_i(q^2).
\end{equation}
Introduction of the separable expansion clearly
breaks translational invariance and  must be regarded purely as a numerical
procedure,
much like a lattice breaks translation invariance. Translation invariance is
restored
as the rank of the separability is increased. The infrared hadronic region
appears to be
well described by a rank $n=2$ form for $D_0$.

The DSE then have solutions of the form
\begin{equation}
B(s)=\sum b_i\Gamma_i(s), ...
\end{equation}
 Equations for the
 $b_i,..$ are easily determined by substituting these forms back into (5) and
(6), giving coupled
transcendental non-linear equations easily solved by iteration.
Then  $\sigma_s$ and $\sigma_v$ are  seen to have the form of sums
\begin{equation}
\sigma_s(s)=\sum_{i=1,n}\sigma_s(s)_i, \mbox{ \ \ \ \ }
\sigma_v(s)=\sum_{i=1,k}\sigma_v(s)_i ,
\end{equation}

However this, in principle, procedure suffers from the defect that since $D$ is
unknown
then some ad hoc choice of the parametrised $\Gamma_i$ must be made. The
resulting
$\sigma_s(s)_i$ and $\sigma_v(s)_i$ develop spurious singularities which impede
the use of
the quark propagator in meson and baryon computations.

A much more robust and physically sensible procedure is to specify a
parametrised form
for the  $\sigma_s(s)_i$ and $\sigma_v(s)_i$ of the chiral-limit quark
propagator with known
analyticity properties. We then  use the DSE in an inverse manner to compute
the
$\Gamma_i(s)$. By using entire functions the much speculated but as yet
unproven quark
confinement property can be ensured. Suitable forms are based on simple model
solutions of the
DSE \cite{entire}. These forms are \begin{equation}
\sigma_s(s)_i=c_iexp(-d_is), \mbox{ \ \ \ \ }
\sigma_v(s)=  \frac{2s-\beta^2(1-exp(-2s/\beta^2))}{2s^2} ,
\end{equation}
where  $\sigma_v(s)$ has only a rank $k=1$ expansion here.

 From these parametrised forms  we can
easily determine the various  parameters $\{b_i,..\}$, which will depend on the
basic, $m=0$,
chiral-limit parameter set ${\cal P}_0=\{c_1,c_2,d_1,d_2,\beta\}$. For example
from (5) we
obtain
\begin{equation}
 b_i^2=\frac{16}{3}\pi^2\int_0^{\infty} sds B(s)_i\frac{B(s)}{sA(s)^2+B(s)^2}
\end{equation}
in which $B(s)=B(s)_1+B(s)_2$, and
\begin{equation}
B(s)_i=\sigma_s(s)_i/(s\sigma_v(s)^2+\sigma_s(s)^2).
\end{equation}

The basic procedure is to find those values of these  parameters ${\cal P}_0$
which allow the
fitting of some small set of hadronic observables to the  experimental values.
We have chosen
$f_{\pi}$ (see \cite{Pi94} for the expression), which  probes exclusively the
space-like region of
the quark propagators, and  the
$a_1$ meson mass which extends the probe  into  the time-like region. The $a_1$
meson is chosen
because all  two-quark states considered here, except for the $1^-$ diquark
correlation, have
lower mass. Thus the region in the complex s-plane where the quark propagator
is needed has
been probed by our fitting procedure. A  space-like only fitting procedure
would require
unreliable and untested extrapolations into the time-like region when computing
other hadronic
observables. The $\pi$ mass is
needed to mainly determine the averaged u  and d current masses.

The $a_1$ and $\pi$ masses are
computed using the Bethe-Salpeter equations (BSE),  keeping only the dominant
Lorentz amplitude
\cite{CRP}.   The BSE are solved with
 the loop integration  in the Euclidean metric, while the meson momentum, in
the rest
frame $P=(\vec{0},iM)$, is in the time-like region of the Minkowski metric.
This
mixed metric is typical of hadron calculations in the GCM, and ensures that the
quark propagators
stay as close as possible to the real positive s-axis, which is the region most
studied for the
quark and gluon propagators. As is well known a separable expansion for the
kernel of an
integral equation, here the gluon propagator in the BSE, reduces that equation
to algebraic form.
The solution of the BSE is thus rendered almost trivial.

We now consider the inclusion of quark current masses in the computations. For
any set of
 ${\cal P}_0$ values, which
implicitly parametrise the   gluon propagator,  we can compute the  non-chiral
quark
propagator by returning to solve (5) and (6) with $m \neq 0$.  The resulting
$\sigma_s$ and
$\sigma_v$ now depend on $m$.  However the approximate DSE is only robust in
the space-like
region. Hence we have fitted, in this region, $\sigma_s(s;m)$ and
$\sigma_v(s;m)$ to the forms
in (13). This ensures that the confinement ansatz continues to hold.
It is this non-chiral quark  propagator which is needed in the $\pi$ and $a_1$
mass
computations. The value of the average u/d quark current mass $m$ is varied,
along with the
${\cal P}_0$ values,  during the fitting procedure. Once the ${\cal P}_0$
values  were known
we were able to use  the DSE  for  the strange quark. Various values of the
strange quark
current mass were used until the K-meson BSE produced the experimental K meson
mass.

The determined parameter values ${\cal P}_0$ are shown in Table 1, and the
values of the fitted
observables are shown in Table 2.  These  three
observables are sufficient to give a  robust determination of the parameter set
and the u/d quark
mass. Table 2 also shows some of the  predicted observables which test the {\em
universality} of
the quark and gluon propagators.  Figs. 1 and 2 show the form of the
chiral-limit $\sigma_s$ and
$\sigma_v$.

Now we consider the  task of reconstructing the translation invariant form for
the
effective gluon propagator. We can easily do this using $D(p)=D_0(p^2,0)$ or
(7) for $D(x)$.
 Hence if the translation invariant constituent quark propagator is known, and
the
separability technique does not compromise that property,  then the full $D(p)$
or its Fourier
transform $D(x)$, determined finally by the values of the parameter set ${\cal
P}_0$, may be
constructed.  It is this $D(p)$, but with the GCM analysis done in the
appropriate gauge, which
should be compared with future direct computations of the gluon propagator.
Using the separable
technique an explicit expression for $D(p)$ may be obtained. We have
$$
D(p)=D_0(p^2,0)
=\sum_i\Gamma_i(p^2)\Gamma_i(0)
=\sum_i\frac{1}{b_i^2}B(p^2)_iB(0)_i
$$
\begin{equation}
=\sum_i\frac{1}{b_i^2}\frac{\sigma_s(0)_i}{\sigma_s(0)^2}\frac{\sigma_s(p^2)_i}
 {p^2\sigma_v(p^2)^2+\sigma_s(p^2)^2},
\end{equation}
an expression which becomes increasingly more accurate as the rank of the
separable expansion is
increased. With the parameter set in Table 1, (14) gives $b_1=0.02672~GeV^2$
and
$b_2=0.02395~GeV^2$, and the resulting $D(p)$ is shown in Fig.3. Note that the
meson data
mandates that the gluon propagator has a  strong IR component (corresponding to
a large
distance confining effect) and a longer range component (corresponding to a
medium distance
effect), as also seen in the QCD studies in \cite{Pennington}.

The predicted hadronic observables in Table 2 are mainly self-evident. The
details of the GCM
computational techniques have all been discussed in the literature. This set of
hadronic  predictions is  obtained without cutoffs or renormalisation
procedures. The meson and
diquark masses are from BSE computations, while the nucleon-core mass
(equivalent to the quenched
approximation in lattice QCD) is from a covariant Faddeev computation
\cite{RTC,Burden} keeping
only the $0^+$ diquark correlation. The separable expansion for the gluon
propagator leads to a
separable form for the diquark correlation propagator.  The   original
nucleon-core Faddeev
computations \cite{Burden} assumed such a separable form, see \cite{RTC} for
details.

Later
computations of the nucleon core \cite{Buck,Ishii,Huang,Meyer} usually had free
parameters that
are adjusted to give the experimental nucleon mass. A full computation of the
nucleon mass in the
GCM requires the inclusion of the $1^+$ diquark correlation, and the dressing
of this nucleon
core by mesons (equivalent to including quark loops in lattice models).

The constituent quark
masses arise from the most probable value of the quark running masses
$(B(s,m)+m)/A(s)$ in the
BSE kernel integrations, which is at $s \approx 0.3$GeV$^2$. Expressions for
the $\pi-\pi$
scattering lengths and the pion charge radius $r_{\pi}$ (without re-scattering
corrections) are
given in \cite{Pi94}.  In \cite{CR85} the MIT bag phenomenology was derived
from the GCM using a
mean field soliton approximation. This gave an expression for the  MIT bag
constant in terms of
$B(s)$ and $A(s)$.

The good values of the predicted observables suggest that the GCM is capable of
 accurately
describing low energy QCD, and that the gluon  propagator multi-rank  separable
expansion  is a
particularly useful and robust computational tool. By including more
experimental data in the
fitted observables   larger rank expansions could be used. This would result in
a more accurate
 mapping out of the gluon and constituent quark propagators in the momentum
range
appropriate to low energy hadronic physics.

RTC acknowledges useful conversations with C.J. Burden, C.D. Roberts, P.C.
Tandy and M.
Thomson.

{\em Addendum}:  Frank and Roberts \cite{Frank95} have reported model gluon
propagator results
also from a fit to meson data using the separable expansion technique.
Burden et al. \cite{Burden95} have also used the separable expansion technique
for the gluon
propagator, but in particular they have studied the meson BSE equation beyond
the dominant Lorentz
amplitude approximation. The results of both works should be compared with the
results reported
here.

\newpage

\vspace{10mm}

\noindent Figure 1:  Plot of extracted $\sigma_s(p)$ $(GeV^{-1})$ plotted
against
                     $p^2$ $(GeV^2)$.

\noindent Figure 2:  Plot of extracted $\sigma_v(p)$ plotted against
                     $p^2$ $(GeV^2)$.

\noindent Figure 3:  Plot of extracted $D(p)$ $(GeV^{-2})$ plotted against
                     $p^2$ $(GeV^2)$.

\vspace{40mm}

\begin{tabular}{llll}
\multicolumn{4}{l} {\bf Table 1: Quark Propagator Parameters: ${\cal P}_0$}\\
\hline \hline
$\mbox{ \ \ }$$c_1$  & 0.5200GeV$^{-1}$ $\mbox{ \ \ \ \ \ \ \ \ \ \ \ \ \ \ \ \
\ }$ &
$c_2$ &1.1794GeV$^{-1}$   \\
$\mbox{ \ \ }$$d_1$  & 2.0737GeV$^{-2}$ & $d_2$    & 4.7214GeV$^{-2}$\\
$\mbox{ \ \ }\beta$    & 0.5082GeV  &    \\
\hline

\end{tabular}

\begin{tabular}{l r r}
\multicolumn{3} {l}
{\bf Table 2:  Hadronic Observables}\\
\hline \hline
 {\bf Observable} & {\bf Theory} & {\bf Expt./Theory[Ref]} \\ \hline \hline
{\bf Fitted observables}\\

$f_{\pi}$  &    93.00MeV  &93.00MeV\\
$a_1$ meson mass  & 1230MeV & 1230MeV\\
$\pi$ meson mass   &  138.5MeV   & 138.5MeV\\
$K $ meson mass (for $m_s$ only)   &  496MeV      &     496MeV\\
\hline
{\bf Predicted observables}\\
$(m_u+m_d)/2$   &  6.5MeV  &  6.0\cite{Bijnens}\\
$m_s$   &  135MeV   &  130MeV\cite{Cheng}\\
$\omega$ meson mass &  804MeV  &  782MeV  \\
$a^0_0$ $\pi-\pi$ scattering length &  0.1634  &  $0.21 \pm 0.01$ \\
$a^2_0$ $\pi-\pi$ scattering length&  -0.0466  &  $-0.040 \pm 0.003$ \\
$a^1_1$  $\pi-\pi$ scattering length&  0.0358   &  $0.038 \pm 0.003$ \\
$a^0_2$ $\pi-\pi$ scattering length&  0.0017   & $0.0017 \pm 0.003$\\
$a^2_2$ $\pi-\pi$ scattering length &  -0.0005  & not measured\\
$r_{\pi}$ pion charge radius & 0.55fm &  0.66fm\\
nucleon-core mass  &  1390MeV &  $\sim1300$MeV\cite{Thomas91}  \\
constituent quark rms size   & 0.59fm& -\\
chiral quark constituent mass &   270MeV&   -\\
u/d quark constituent  mass &  300MeV  & $\sim 340$MeV\cite{Halzen}\\
s quark constituent  mass &  525MeV  & $\sim510$MeV\cite{Halzen}\\
$0^+$ diquark rms size &   0.78fm   & -\\
$0^+$ diquark constituent mass &  692MeV  &  $>$400MeV  \\
$1^+$ diquark constituent  mass &  1022MeV  &  -  \\
$0^-$ diquark constituent  mass &  1079MeV  &  -  \\
$1^-$ diquark constituent  mass &  1369MeV  &  -  \\
MIT bag constant  &  (154MeV)$^4$  & (146 MeV)$^4$ \\
MIT nucleon-core mass (no cm corr.)  & 1500MeV &   $\sim$1300MeV   \\
\hline
\end{tabular}


\newpage
\begin{thebibliography}{99}

\bibitem{CR85} R.T. Cahill and C.D. Roberts, Phys. Rev. D {\bf 32} (1985)2419.

\bibitem{Pennington} N. Brown and M.R. Pennington, Phys. Rev. D {\bf
39}(1989)2723.


\bibitem{RTC} R.T. Cahill, Aust. J. Phys. {\bf 42}(1989)171; Nucl. Phys. A {\bf
543}(1992)63c.

\bibitem{Reinhardt90} H. Reinhardt, Phys. Lett. B {\bf 244}(1990)316.

\bibitem{Frank} M.R. Frank and P C. Tandy, Phys. Rev. C {\bf 46}(1992)338.

\bibitem{Stainsby94} S.J. Stainsby and R.T.Cahill, Mod. Phys. Lett. A {\bf
9}(1994)3551.

\bibitem{entire} C.J. Burden, C.D. Roberts and A.G. Williams, Phys. Lett. B
{\bf 285}(1992)347.

\bibitem{Pi94}  C.D. Roberts, R.T. Cahill, M.E. Sevior and N. Iannella,
                Phys. Rev. D {\bf 49}(1994)125.

\bibitem{CRP} R.T. Cahill, C.D. Roberts and J. Praschifka, Phys. Rev. D {\bf
36}(1987)2804.


\bibitem{Burden} C.J. Burden, R.T. Cahill and J. Praschifka, Aust. J. Phys.
{\bf 42}(1989)147.

\bibitem{Buck} A. Buck, R. Alkofer and H. Reinhardt, Phys. Lett. B {\bf
286}(1992)29.

\bibitem{Ishii} N. Ishii, W. Bentz, and K. Yazaki, Phys. Lett. B {\bf
301}(1993)165;
                Phys. Lett. B {\bf 318}(1993)26.

\bibitem{Huang} S. Huang and J. Tjon, Phys. Rev. C {\bf 49}(1994)1702.

\bibitem{Meyer} H. Meyer, Phys. Lett. B {\bf 337}(1994)37.


\bibitem{Bijnens} J. Bijnens, J. Prades and E. de Rafael, Phys. Lett. B {\bf
348}(1995)226.

\bibitem{Cheng} T-P Cheng and L-F Li, {\em Gauge Theory of Elementary
Particles}, Oxford 1988.

\bibitem{Thomas91} A.W. Thomas, Aust. J. Phys. {\bf 44}(1991)173.


\bibitem{Halzen} F. Halzen and A.D. Martin, {\em Quarks and Leptons}, Wiley
1984.

\bibitem{Frank95} M.R. Frank and C. D. Roberts, hep-ph/9508225.

\bibitem{Burden95} C.J. Burden, L. Qian, C.D. Roberts, P.C. Tandy and M.Thomson
 ANU /Argonne /Kent State 1995 report, in preparation.





\end{thebibliography}
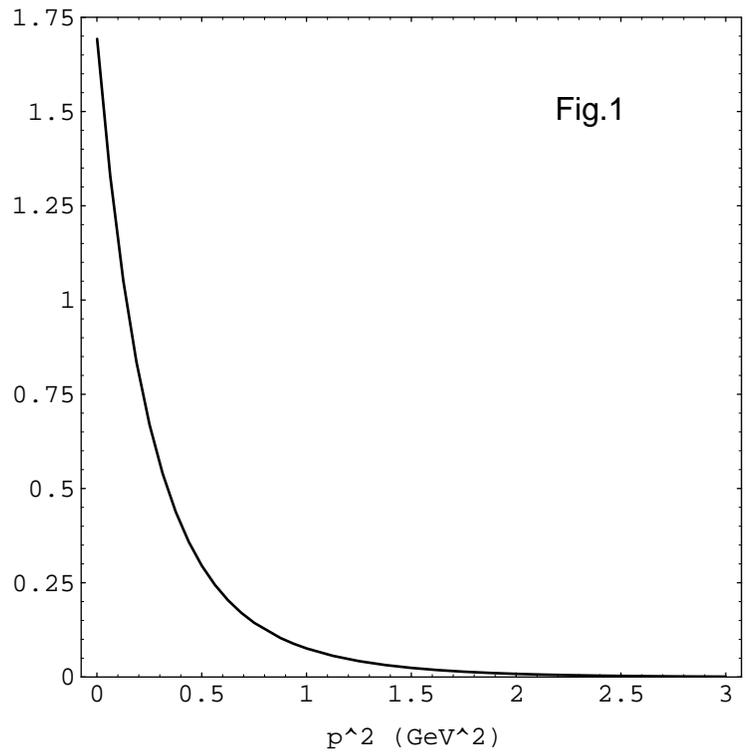
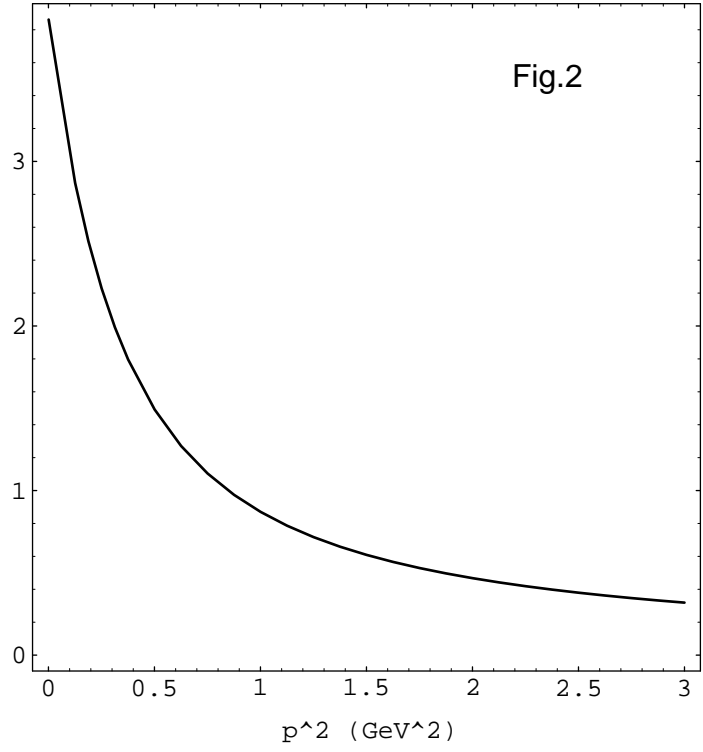
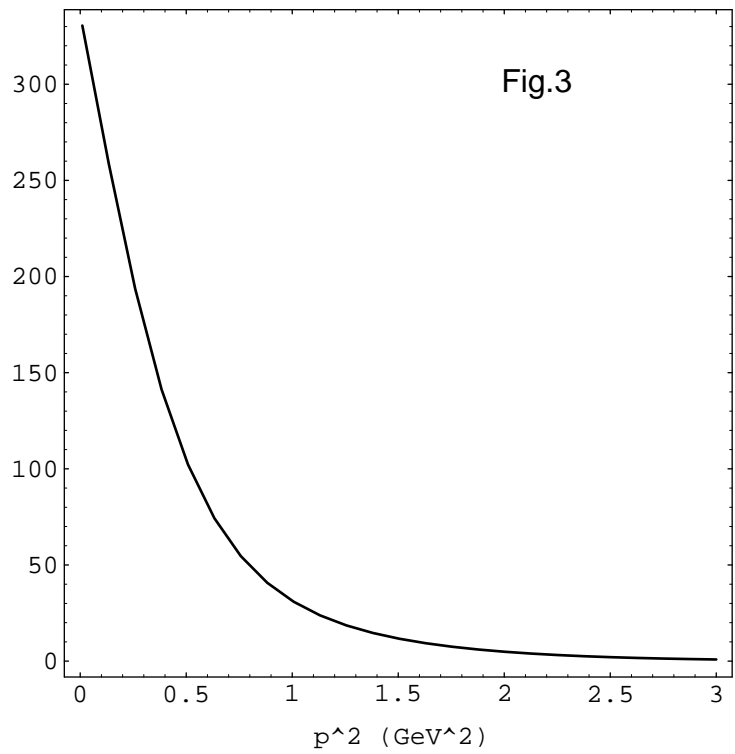
\end{document}